 \documentstyle[aps,prl,twocolumn,epsfig]{revtex}

\begin{document}
\title{Macroscopic dynamics of a trapped Bose-Einstein 
condensate in the presence of 1D and 2D optical lattices}
\author{M. Kr\"amer$^{a}$, L. Pitaevskii$^{a,b}$ and S. Stringari$^{a}$}
\address{$^{a}$Dipartimento di Fisica, Universit\`{a} di Trento,}
\address{and Istituto Nazionale per la Fisica della Materia,}
\address{I-38050 Povo, Italy}
\address{$^{b}$ Kapitza Institute for Physical Problems,} 
\address{ ul. Kosygina 2, 117334 Moscow, Russia}
\date{\today}
\maketitle

\begin{abstract}
The hydrodynamic equations of superfluids 
for a weakly interacting Bose gas
are generalized to include
the effects of periodic optical potentials produced by 
stationary laser beams.
The new equations are 
characterized by a renormalized interaction coupling constant
and by an effective mass accounting for the inertia of the system along the
laser direction. 
For large laser intensities the effective mass is directly
related to the tunneling rate between two consecutive wells. 
The predictions for the  frequencies of the collective modes
of a condensate confined by a magnetic harmonic trap are discussed for both 1D and 2D optical lattices and 
compared with recent experimental data.

\end{abstract}

\bigskip

\narrowtext

The  experimental realization of  optical lattices 
 \cite{kas1,kas2,florence,munich1,arimondo,munich2} 
is stimulating new perpectives
in the study of coherence phenomena in trapped Bose-Einstein condensates. 
A first direct measurement
of the critical Josephson current has been recently obtained  
in \cite{florence} 
by studying the center of
mass motion of a magnetically trapped gas 
in the presence of a 1D periodic optical potential. 
Under these conditions the propagation of  collective modes
is a genuine quantum effect produced by the
tunneling through the barriers  and by the 
superfluid behaviour associated with 
 the coherence of the order parameter
between different wells. The effect
of the optical potential is to increase the inertia of the gas along the 
direction
of the laser giving rise to a reduction of the frequency of the 
oscillation. 

The purpose of the present work is to investigate the  collective oscillations 
of a magnetically trapped gas in the presence of 1D and 2D optical lattices 
taking into
account  the effect of tunneling,   
 the role of the mean field interaction  and the 3D nature of the sample. 
Under suitable conditions these effects can be described 
by properly generalizing the hydrodynamic
 equations of superfluids  \cite{ss96}.
 
Let us assume that the gas, at $T=0$, be trapped by an external potential given by the
sum of  a 
harmonic trap of magnetic origin $V_{\rm ho}$ and of a stationary
 optical potential $V_{\rm opt}$ modulated along the $z$-axis. 
 The resulting potential is given by 
  \begin{equation}
 V_{ext} = {1\over 2} m \left(\omega^2_xx^2 + \omega_y^2y^2 + \omega_z^2z^2
 \right) + s E_R\sin^2qz 
 \label{V}
 \end{equation}
 where $\omega_x$, $\omega_y$, $\omega_z$ are the frequencies of the harmonic  
trap, 
 $q = 2\pi /\lambda$ is fixed by the wavelength of the laser light creating
 the stationary 1D lattice 
 wave,   $E_R  = \hbar^2q^2/2m$ is the so called recoil energy and $s$ is a 
 dimensionless parameter providing  the intensity of the 
 laser beam. 
 The optical potential has periodicity 
 $d=\pi /q =\lambda/2$  along the $z$-axis. 
The case of a 2D lattice will be discussed later.
In the following we will assume that the laser intensity be large enough
 to create many separated wells giving rise to an array of several condensates.
 Still, due to quantum tunneling, the overlap between the wave functions of two consecutive wells
 can be sufficient to ensure full coherence. In this case 
 one is allowed to use the Gross-Pitaevskii (GP) theory for the 
order parameter to study both the equilibrium and the
 dynamic behaviour of the system at zero temperature \cite{RMP}. 
Eventually, 
 if the tunnelling  becomes 
 too small, the fluctuations of the relative  phase 
between the condensates will destroy
 the coherence of the sample giving rise to new quantum configurations associated
 with the transition to a Mott insulator phase \cite{kas2,munich2}.   
 
In the presence of coherence it is natural to  make the ansatz 
 \begin{equation}
 \Psi({\bf r}) = \sum_k \Psi_k(x,y)f_k(z) e^{iS_k(x,y)}
 \label{Psi}
 \end{equation}
 for the order parameter in terms of a sum of many condensate
 wave-functions relative to each well. 
Here $S_k(x,y)$ is the phase of the $k$-component 
 of the order parameter, while $\Psi_k$ and $f_k$ are real functions. 
  We will make the further periodicity assumption 
 $f_k(z)= f_0(z-kd)$ 
where  $f_0$ is localized at the origin. 
    The above  assumptions for $\Psi$ and $f_k$  are 
 justified for relatively large values of $s$ 
where the interwell barriers 
are significantly higher than the chemical potential. 
In this case 
 the condensate wave functions of different sites are well separated 
(tight binding approximation).

Using the ansatz (\ref{Psi}) for the order parameter one finds the 
following result 
for  the mean field expectation value
of the effective Hamiltonian
$H = \sum_j \left({\bf p}^2_j/2m + V_{ext}({\bf r}_j)\right) + g\sum_{j < k}\delta
({\bf r}_j-{\bf r}_k)$:
\begin{eqnarray}
&\phantom{=}&E=\langle H \rangle=
\left[\int\!dz\,\frac{\hbar^2}{2m}\left( \partial_z f_0\right)^2+
f_0^2 V_{\rm opt}\right]\sum_k\!\int\!\!dxdy\, \Psi_k^2
\nonumber\\
&+&
\frac{g}{2}\,\left[\int\!dz f_0^4\right]
\sum_k\!\int\!\!dxdy\, \Psi_k^4 
\nonumber\\
&+&
\left[\int\! dz f_0^2\right]
\sum_k \!\int \!\!dxdy 
\left[\frac{\hbar^2}{2m}\left(\partial_{{\mathbf r}_{\perp}} 
\Psi_k\right)^2\right. 
\nonumber\\
&\phantom{=}&\phantom{=}
+\left.\Psi_k^2 V_{\rm ho}({\mathbf r}_{\perp},kd)
+\frac{\hbar^2}{2m}\Psi_k^2 
\left(\partial_{{\mathbf r}_{\perp}} S_k\right)^2\right]
\nonumber\\
&-&\delta
\sum_k\!\int \!\!dxdy \,\Psi_k\Psi_{k+1}\,{\rm cos}\left[S_k-S_{k+1}\right]\,,
\label{E}
\end{eqnarray}
where in the two-body and in the magnetic interaction terms as well as 
in the radial kinetic energy
we have ignored the overlap contributions arising from different wells. 
In the evaluation of the axial kinetic energy and of the optical potential term we have instead
kept also the overlap terms originating from  consecutive wells. 
These are proportional to the quantity
\begin{equation}
\delta=-2\!\!\int \!dz
{\hbar^2\over 2m}\partial_z f_0(z) \partial_z f_0(z\!-\!d)\!+\!f_0(z) f_0(z\!-\!d)
V_{\rm opt}\,,
\label{delta}
\end{equation}
related to the tunneling rate and  
responsible for the occurrence of Josephson effects.

By setting $S_k=0$ (groundstate configuration),
the variation of $E$ with respect to $f_0$ 
yields the differential
equation 
\begin{equation}
\left[-{\hbar^2 \over 2m} {\partial^2 \over \partial z^2}  
+ s E_R \sin^2qz\right]f_0(z) = \varepsilon_0
f_0(z)
\label{eqf}
\end{equation}
where $\varepsilon_0$ is introduced to ensure the normalization
condition $\int_{-d/2}^{d/2} dz f_0^2=1$ 
which implies that the functions $\Psi_k$ are 
normalized to the number of atoms $N_k$ occupying each site: 
$\int \Psi_k^2 dxdy=N_k$. 
In eq. (\ref{eqf}) we have ignored the contribution arising from the 
two-body interaction. Estimates of \cite{pedri} show that this is a good 
approximation already at moderately large $s$. 
We have also neglected the external 
magnetic potential which is justified if $\sqrt{s}E_R>>\hbar\omega_x, \hbar\omega_y$.
Since in the following we are interested in the low energy
excitations of the system we 
will always keep the function $f_0$ equal 
to the groundstate solution of (\ref{eqf}).

In order to discuss the macroscopic properties of the system, including
its
low energy dynamics,  it is convenient
to transform the discretized formalism described above into the one of 
continuum variables.
This is obtained through the replacement $\sum_k \to (1/d)\int\!dz$ 
in the various
terms of the energy.
Through such a procedure one naturally introduces a smoothed or  
"macroscopic"  density defined by 
\begin{equation}
n_{M}(x,y,z) = (1/d) \Psi_k^2(x,y)
\label{nM}
\end{equation}
with $z \simeq d\,k$, and a smoothed phase $S$ 
($S_k\rightarrow S(x,y,z)$).

By applying the  smoothing procedure to eq.(\ref{E})
we obtain the following macroscopic expression for the energy functional
\begin{eqnarray}
E=\!
\int\! dVn_M\!\left[\frac{\tilde{g}n_M}{2}+V_{\rm ho}
+\frac{\hbar^2}{2m}\!\left(\partial_{{\mathbf r}_{\perp}} 
S\right)^2-
\delta {\rm cos}\left[d\partial_z S\right]\right]
\,,
\label{AMacro}
\end{eqnarray}
where we have introduced the renormalized coupling constant
$
\tilde{g}=gd\int f_0^4dz\,,
$
we have
neglected quantum pressure terms originating from the radial term in the
kinetic energy
 and
we have set $\Psi_k \Psi_{k+1} \sim \Psi_k^2=d n_M$.
We have also omitted some constant terms (first two terms in eq. (\ref{E}))
 which do not depend on $n_M$ or on $S$.

With respect to the functional characterizing a trapped Bose gas in the absence of optical confinement, one notices two important differences:
first the interaction coupling constant is renormalized due to 
the presence of the  optical lattice.
This is the result of the local compression of the gas produced by the tight
optical confinement which increases the repulsive effect of the interactions.
Second the kinetic energy term along the $z$-direction has no longer the classical 
quadratic form
 as in the radial direction, but exhibits a periodic 
dependence on the 
 gradient of the phase. 
 By expanding this term for small gradients, 
which is the case in the study of small amplitude oscillations, 
one derives a quadratic term of the form $(\hbar^2/2m^*)\int dV n_M(\partial_z S)^2$
characterized by the effective mass 
\begin{equation}
{m \over m^*} = {m\delta d^2\over \hbar^2}={\delta\over E_R}{\pi^2\over 2}
\label{mstar}
\end{equation}
where $\delta$ is defined by eq. (\ref{delta}). Notice that within the employed approximation
the value of $\delta$, and hence of $m^*$, does not depend on the number of atoms, nor on
the mean field interaction.

The equilibrium density profile, obtained by minimizing eq.(\ref{AMacro}) with 
$S=0$ has the typical form of an inverted parabola \cite{notepedri}
\begin{equation}
n_{M}^0 = \left(\mu -{1\over 2}m (\omega^2_xx^2 + 
\omega_y^2y^2 + \omega_zz^2)\right)/\tilde{g}\,,
\label{nMeq}
\end{equation}
which conserves the aspect ratio of the original magnetic trapping. 
The size of the condensate has instead increased since $\tilde{g}>g$.
For large $s$
the increase of the coupling constant can be large 
($\tilde{g}\sim s^{1/4}$ 
\cite{pedri}). 
However, 
since the radius of the sample scales like the 1/5-th power  of $\tilde{g}$ 
the resulting increase in the size of the system is not very spectacular 
(for $s=15$ we find an increase of the size by $\sim 20\%$ for the 
experimental setting of \cite{florence}).

 The functional (\ref{AMacro}) can be used to carry out dynamic calculations.
 In this case one needs the action
$
A = \int dt\left( \langle H \rangle - 
i\hbar \langle {\partial \over \partial t}
\rangle \right)\,,
$
with the second term given by  
 $i\hbar\langle(\partial/ \partial t)\rangle = -
\int dV \hbar n_M \dot{S}
$.
The resulting equations of motion are obtained by imposing the
stationarity condition on the action
with respect to arbitrary variations
of  the density $n_M$
and of the phase $S$. 
The equations take the form
\begin{eqnarray}
\dot{n}_M+
{\hbar\over m}{\partial_{{\mathbf r}_{\perp}}}
\left(n_M{\partial_{{\mathbf r}_{\perp}}}S\right)
+{\delta d\over\hbar}{\partial_z}
\left(n_M {\rm sin}\left[d{\partial_z}S\right]\right)
&=&0\,,
\label{eqsm1}
\\
\hbar\dot{S}+\tilde{g}n_M+V_{mag}+
{\hbar^2\over 2m}\left(\partial_{{\mathbf r}_{\perp}}S\right)^2
-\delta {\rm cos}\left[d\partial_z S\right]
&=&0\,.
\label{eqsm}
\end{eqnarray}
In particular, at equilibrium 
these equations reproduce result (\ref{nMeq}) for the 
equilibrium density.
Furthermore, Josephson-type oscillations are among those captured by 
eqs. (\ref{eqsm1}) and (\ref{eqsm}).
To see this consider the case of a uniform gradient of the phase  along $z$, 
$\partial_z S={P_Z(t)/\hbar}$, where $P_Z$ is a 
time-dependent parameter.
From eqs. (\ref{eqsm1}) and (\ref{eqsm}) one can then derive
equations of motion for the 
center of mass 
$Z(t)=\int dV z n_m(t)/N$ and for the conjugate momentum variable $P_Z$
\cite{florence,andrea}
\begin{eqnarray}
\hbar\dot{Z}-\delta d {\rm sin}\left[d{P_Z\over\hbar}\right]&=&0\,,
\label{cm2a}\\
\dot{P}_Z+m\omega_z^2 Z&=&0
\,,
\label{cm2}
\end{eqnarray}
which have the typical Josephson form.

In the limit of small oscillations the solutions of 
eqs. (\ref{eqsm1}) and (\ref{eqsm}) have  
the form $n=n_M^0+\delta n({\mathbf r})e^{i\omega t}$ with 
$\delta n$ obeying 
the 
hydrodynamic equations:
\begin{eqnarray}
-\omega^2\delta n
&=&
\partial_{{\bf r}_{\perp}}\!\!
\left[
{\mu-V_{\rm ho}\over m}
\partial_{{\bf r}_{\perp}}\delta n
\right]
+
\partial_z\!\!\left[
{\mu-V_{\rm ho}\over m^*}
\partial_z\delta n
\right]
\,,
\label{hdnew}
\end{eqnarray}
where $\mu=\tilde{g}n_M^0(0)$ is the chemical potential of the sample and 
$n_M^0(0)$ is the equilibrium density 
(\ref{nMeq}) evaluated at the center.
The solutions of (\ref{hdnew}) provide the low energy  excitations of the system. In the absence
of magnetic trapping one finds phonons propagating at the velocity 
$c = \sqrt{\tilde{g}n_M^0/m^*}$, in agreement with the result obtained in
\cite{java} for a 1D array of Josephson junctions. In the presence of harmonic trapping 
the discretized frequencies of the time-dependent solutions 
of (\ref{hdnew}) do not
depend on the
value of the coupling constant.
By applying the transformation $z\rightarrow\sqrt{m^*/m}z$, 
one actually finds that the new frequencies are 
simply obtained from  the results of \cite{ss96} by replacing 
\begin{equation}
\omega_z\rightarrow
\omega_z\sqrt{m/m^*}\,.
\end{equation} 
For an elongated trap ($\omega_x=\omega_y=\omega_{\perp} \gg \omega_z$) 
the lowest solutions
are given by the center-of-mass motion $\omega_D = \sqrt{m/m^*}\omega_z $ and by the quadrupole 
mode $\omega_Q = \sqrt{5/2}\sqrt{m/m^*}\omega_z $. 
The center-of-mass frequency coincides with the value obtained from eqs. 
(\ref{cm2a}) and (\ref{cm2}) in the limit of small oscillations. 
Concerning the quadrupole frequency we note 
that the occurrence of the factor $\sqrt{5/2}$ is a non-trivial consequence of the mean field interaction predicted by the hydrodynamic theory of superfluids in the presence of harmonic
trapping \cite{ss96}.
In addition to the low-lying axial motion the system exhibits radial 
oscillations at high frequency, of the order of $\omega_{\perp}$. 
The most important
ones are the transverse breathing and quadrupole oscillations occuring at 
$\omega=2\omega_{\perp}$ and $\omega=\sqrt{2}\omega_{\perp}$
respectively.
For elongated traps the frequencies of these 
modes should not be affected by the presence of the optical
potential.
Different scenarios are obtained for disc-shaped traps 
($\omega_z>>\omega_{\perp}$).
The above results apply to the linear regime of small oscillations.
Eqs. (\ref{cm2a}) and (\ref{cm2}) show that in the case of 
center-of-mass oscillations, the
linearity condition is achieved for initial displacements $\Delta x$ of the
trap satisfying $\Delta x<<\sqrt{2\delta/m\omega_z^2}$, 
a condition that becomes more and more severe as the laser intensity increases.
For larger initial displacements the oscillation is 
described by the pendulum equations. For very large amplitudes 
the motion is however dynamically 
unstable \cite{andrea,smerzi-gp-dipole}.

\begin{figure}
\includegraphics[width=8cm,height=5cm]{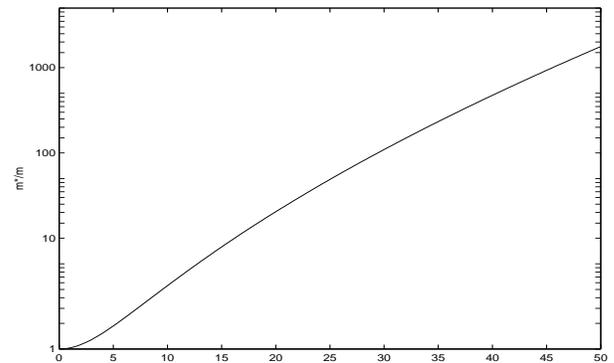}
\caption{
Effective mass 
as a function of the laser intensity $s$ 
(see eq.(\ref{V})) calculated neglecting the effects of interaction and 
harmonic trapping.
}
\label{fig-m-mstar}
\end{figure}

From the previous discussion it emerges that the effective mass is the crucial
parameter needed to
predict the value of the small amplitude collective frequencies. 
An estimate
of $m/m^*$ can
be made by neglecting the magnetic trapping as well as the role 
of the mean field
interaction. 
Within this approximation
the effective mass is easily obtained from 
the excitation spectrum of the Schr\"odinger equation for the 1D Hamiltonian 
$H = -(\hbar^2/2m) \partial^2/\partial z^2 +sE_R\sin^2qz $, 
avoiding the explicit determination of the tunneling parameter (\ref{delta}). 
One looks for
solutions of the form $ e^{ipz/\hbar}f_{p}(z)$ where 
$p$ is the quasi-momentum of the atom and $f_{p}(z)$ is a
 periodic function 
of period $d$.
 The resulting dispersion law $\varepsilon(p)$
 provides, for small ${p}$, the effective mass according to
 the identification $\varepsilon(p)  \simeq \varepsilon_0 + 
 p^2/2m^*$.
 The value of $m/m^*$, which turns out to be a universal function of the intensity
parameter $s$, has been evaluated for
a wide range of values of $s$ (see fig.\ref{fig-m-mstar}).  
These results for $m^*$ 
can be used to estimate 
the actual value of the collective frequencies. 
The method described here to calculate $m^*$ is expected to be reliable not 
only for very large laser intensities $s$ when the tight binding approximation applies and the effective mass can be expressed in terms of the tunneling rate (see eqs. (\ref{mstar}),(\ref{delta})), but also for smaller values of $s$. 
Of course for very small laser intensities, as in the experiment 
\cite{florence-prl}, the 
determination of $m^*$ requires the inclusion of the 
mean field interaction and of the magnetic trapping through the explicit 
solution of the GP-equation.

\begin{figure}
\includegraphics[width=8cm,height=5.2cm]{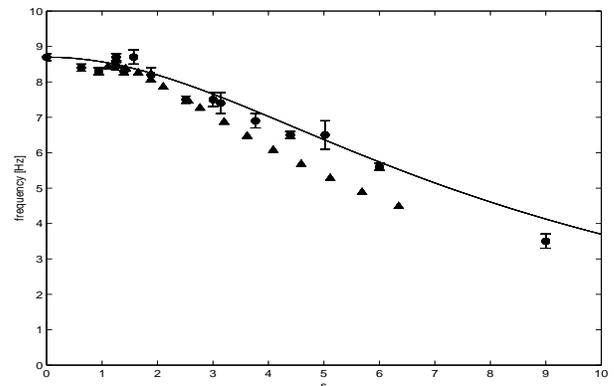}
\caption{Frequency of the center-of-mass motion for a condensate trapped
by the combined magnetic and optical potential (\ref{V}) 
as a function of the laser
intensity.
The circles and triangles are, respectively, 
the experimental and theoretical data of 
{\protect\cite{florence}}.
The triangles have been obtained by
evaluating the tunneling rate within a Gaussian approximation 
for the order parameter in each well {\protect\cite{florence}}. 
The solid line refers to our theoretical prediction.}
\label{fig-dipolefreq}
\end{figure}

In fig. \ref{fig-dipolefreq} we compare our predictions for the frequencies 
of the center-of-mass motion with the recent experimental data obtained in
\cite{florence}.
The comparison reveals good agreement with the experiments.
Our results also agree well with those obtained from the numerical solution of the time-dependent GP-equation \cite{andrea,smerzi-gp-dipole}.

The above formalism is naturally generalized to include a 2D optical 
lattice where the optical potential is  
$V_{\rm opt}=s E_R {\rm sin}^2q x+s E_R {\rm sin}^2q y$.
The actual potential now generates an array of 1D condensates which has 
already been the object of experimental studies \cite{munich1}. 
For a 2D-lattice the  
ansatz for the order parameter is \cite{note}
\begin{equation}
 \Phi({\bf r}) = \sum_{k_x,k_y} \Psi_{k_x,k_y}(z)f_{k_x,k_y}(x,y) 
e^{iS_{k_x,k_y}(z)}\,.
 \label{Psi2D}
\end{equation}
In the TF-limit
the groundstate
smoothed density $n_M=\Psi_{k_x,k_y}^2/d^2$
still has the familiar 
form 
$n_M^0
=
\left(
\mu-V_{\rm ho}
\right)/\tilde{g}$
with the redefined coupling constant 
$\tilde{g}=g\left(d\int\!dx f_0^4\right)^2$, where $f_0$ is still given by 
the solution of eq. (\ref{eqf}) and we have used the same approximations 
as in the 1D case.

Also with regard to dynamics, one can proceed as for the 1D lattice.
One finds that the equations of motions, after
linearization, take the form
\begin{eqnarray}
\ddot{\delta n}
&=&
\partial_z\left[
{\mu-V_{\rm ho}\over m}
\partial_z\delta n
\right]
+
\partial_{{\mathbf r}_{\perp}}
\left[
{\mu-V_{\rm ho}\over m^*}
\partial_{{\mathbf r}_{\perp}}\delta n
\right]
\,.
\label{eqmotion2-lin52Dsum}
\end{eqnarray}
The frequencies of the low energy collective modes are then obtained from those
in the absence of the lattice \cite{ss96} by simply replacing
$\omega_{x}\rightarrow\sqrt{m/ m^*}\omega_{x}$ and 
$\omega_{y}\rightarrow\sqrt{m/ m^*}\omega_{y}$.
For large laser intensities
the value of $m^*$
coincides with the one calculated for the 1D array.
If $\omega_z>>\omega_x\sqrt{m/m^*}, \omega_y\sqrt{m/m^*}$, 
the lowest energy solutions involve the motion in the $x-y$ plane. 
The oscillations in the $z$-direction are instead 
fixed by the value of $\omega_z$. 
These include the center-of-mass motion ($\omega=\omega_z$) and the
lowest compression mode ($\omega=\sqrt{3}\omega_z$) \cite{ss96,RMP}.
The frequency $\omega=\sqrt{3}\omega_z$ coincides with the value obtained by 
directly applying the hydrodynamic theory to 1D systems \cite{ss98,ho}
 and reveals the 1D nature of the tubes generated by the 2D lattice.
If the radial trapping generated by the lattice becomes too strong the motion 
along the tubes can no longer be described by the mean field equations and 
one jumps into more correlated 1D regimes \cite{chiara}.

\bigskip

Stimulating discussions with F. Cataliotti, C. Fort, M. Inguscio, A. Smerzi 
and A. Trombettoni are acknowledged.
This research is supported by the Ministero della Ricerca Scientifica e 
Tecnologica (MURST).



\begin{references}

\bibitem{kas1}  B.P.~Anderson and M.A.~Kasevich,  Science {\bf 282}, 1686
(1998).

\bibitem{kas2} C.\ Orzel, A.K.~Tuchman, M.L.~Fensclau, M.~Yasuda, and
M.A.~Kasevich, Science {\bf 291}, 2386 (2001).

\bibitem{florence}
F.S.~Cataliotti, S.~Burger, C.~Fort, P.~Maddaloni, F.~Minardi, A.~Trombettoni,
 A.~Smerzi, M.~Inguscio, Science {\bf 293}, 843 (2001).

\bibitem{munich1}
M.~Greiner, I.~Bloch, O.~Mandel, T.W.~H\"ansch, T.~Esslinger, Phys. Rev. Lett. {\bf 87}, 160405 (2001)..

\bibitem{arimondo}
O.~Morsch, J.H.~M\"uller, M.~Cristiani, D.~Ciampini, E.~Arimondo, Phys. Rev. Lett. {\bf 87}, 140402 (2001).

\bibitem{munich2} 
M.~Greiner, O.~Mandel, T.~Esslinger, T.W.~H\"ansch,  
I.~Bloch, Nature {\bf 415}, 39 (2002).

\bibitem{ss96} 
S.~Stringari, Phys. Rev. Lett. {\bf 77}, 2360 (1996). 

\bibitem{RMP}  F.~Dalfovo, S.~Giorgini, L.P.~Pitaevskii, and S.~Stringari,
Rev. Mod. Phys. {\bf 71}, 463 (1999).

\bibitem{pedri} P.~Pedri, L.~Pitaevskii, S.~Stringari, C.~Fort, S.~Burger,
F.S.~Cataliotti, P.~Maddaloni, F.~Minardi, M.~Inguscio, Phys. Rev. Lett. 
{\bf 87}, 220401 (2001). 

\bibitem{notepedri}
The profile (\ref{nMeq}) can also be obtained 
by applying the smoothing procedure (\ref{nM}) to the equilibrium solution for 
$\Psi_k^2$ given by 
eq.(8) in \cite{pedri}.

\bibitem{andrea}
A.~Trombettoni, PhD-thesis, SISSA, Trieste (2001).

\bibitem{java}  J.~Javanainen, Phys. Rev. A {\bf 60}, 4902
(1999).


\bibitem{smerzi-gp-dipole}
A.~Trombettoni,
A. Smerzi, unpublished. 

\bibitem{florence-prl}
S.~Burger, F.S.~Cataliotti, C.~Fort, F.~Minardi, M.~Inguscio, M.L.~Chiofalo 
and M.P.~Tosi, Phys. Rev. Lett. {\bf 86}, 4447 (2001).

\bibitem{note}
In analogy with the results of \cite{pedri} we find that, at equilibrium, the 
quantity $\Psi^2_{k_x,k_y}$ is given by an inverted parabola as a function of
$z$. The number of particles $N_{k_x,k_y}=\int dz \Psi^2_{k_x,k_y}$ 
occupying the
corresponding site is given by
$N_{k_x,k_y}
=N_{0,0} \left(1-{k_x^2/K_x^2}-{k_x^2/K_y^2}\right)^{3/2}$
with $K_{x,y}=
\sqrt{\hbar\bar{\omega}/ m\omega_{x,y}^2d^2}
\left(15{N a}\left(d\int\!dxf_0^4\right)^2/ a_{\rm ho}\right)^{1/5}$
and
$N_{0,0}
={5\over 2\pi}{N/ K_x K_y}$. 

\bibitem{ss98}
S.~Stringari, Phys. Rev. A {\bf 58}, 2385 (1998).

\bibitem{ho}
T.-L.~Ho and M.~Ma, J. Low. Temp. Phys. {\bf 115}, 61 (1999).

\bibitem{chiara}
C.~Menotti and S.~Stringari, cond-mat/0201158.

\end{references}
\end{document}